\documentclass[pre,twocolumn,float,showpacs,amsmath,amssymb]{revtex4}
\usepackage{graphicx}
\usepackage{dcolumn}
\usepackage{bm}

\begin{document}

\title
{Universal SSE algorithm for Heisenberg model and Bose Hubbard
model with interaction}

\author{M. V. Zyubin}
\email{mikhail$_$zyubin@yahoo.com}
\author{V. A. Kashurnikov}
\email{kash@pico.mephi.ru}

\address{ Moscow Engineering Physics Institute, 115409 Moscow, Russia}

\date{\today}
\begin{abstract}
We propose universal SSE method for simulation of Heisenberg model
with arbitrary spin and Bose Hubbard model with interaction. We
report on the first calculations of soft-core bosons with
interaction by the SSE method. Moreover we develop a simple
procedure for increase efficiency of the algorithm. From
calculation of integrated autocorrelation times we conclude that
the method is efficient for both models and essentially eliminates
the critical slowing down problem.
\end{abstract}

\pacs{05.10.Ln, 05.30.Jp, 75.10.Jm}

\maketitle

\section{Introduction}
Recently, significant progress in quantum Monte Carlo methods has
been observed. During the last two decades, advanced quantum Monte
Carlo algorithms have been developed. First quantum Monte Carlo
methods, so-called world line algorithms, were based on
Suzuki-Trotter approximation and used local updates
\cite{Hirch,Scal}. It has been replaced by the loop algorithms,
wich use non-local updates. Using of non-local loop updates allows
to decrease autocorrelation times by orders of magnitude
\cite{Evertz}. Later the loop algorithms in continuous imaginary
time have been developed \cite{Wiese}. The continuous-time
implementation of the loop algorithm has eliminated errors
resulting from the Trotter discretization and, hence, loop
algorithms have become numerically exact methods.

Unfortunately, loop algorithms are inefficient in  the presence of
external field \cite{Kaw}. The origin of this slow-down results
from the method of including external field into the simulations.
External field is taken into account through the global weight,
which increases with field increase. To construct efficient
algorithm one should takes into account external field locally, in
the loop construction. For the first time this idea was implemented
in the framework of the worm algorithm \cite{Kash}.

Both worm and loop algorithms work directly in continuous imaginary
time. At the same time there is a numerically exact quantum Monte
Carlo method, which works in the discrete basis. It is Stochastic
Series Expansion (SSE)  method. SSE algorithm is based on power
series expansion of a partition function. Initially SSE method was
developed with local updates \cite{Sand1}. Later the algorithm with
loop updates was proposed \cite{Sand2}. Applying loop updates for
SSE method has same favorable consequence as for world line
algorithms and SSE method has become powerful tool for exploring
quantum many-body systems. Recently Sandvik and Sylju{\aa}sen
introduce the concept of directed loops in stochastic series
expansion, which allows to perform simulation in wide range of
external fields \cite{Sand3}.

Last years loop algorithms and SSE algorithm have been used for
exploring of different quantum systems. Investigations of quantum
spins \cite{sp,Todo,spin1,sps}, bosons \cite{bos} and
one-dimensional fermion systems \cite{ferm} have been performed.
However, at the present moment investigations of hard-core bosons
and spin $S=1/2$ systems are predominant in literature.

The authors of the papers \cite{Todo} investigate spin systems with
spin $S>1/2$ by loop algorithms. But they do not take into account
external field. And they use the spin-split representation, i.e.
replace the original spin operators by the sum of $2S$ Pauli
operators. Such representation is uncomfortably because it requires
extra memory resources and it cannot be applied directly for
soft-core bosons.

P. Henelius et al. have studied ferromagnetic Heisenbegr model with
spin up to $S=2$ in the wide range of external field by using of
the SSE algorithm \cite{sps}. Our calculations indicate that the
standard SSE algorithm is quite effective in the case of
ferromagnetic Heisenberg model but for simulation Heisenberg
antiferromagnet it is necessary to increase efficiency of the
algorithm.

Until now we do not know about simulations of soft-core bosons by
the loop or SSE algorithms. Very recently Kawashima et al. develop
method for {\sl free} soft-core bosons based on the mapping of
bosonic models to the spin models \cite{Smak}. For simulation of
spin system they use coarse-grained loop algorithm with the
spin-split representation. Unfortunately, the authors do not give
any quantitative characteristics  of their algorithm efficiency.

In the present work we propose universal algorithm based on SSE
method, which allows to investigate both spin systems with
arbitrary spin in the presence of external field and systems of
{\sl interacting} soft-core bosons in the presence of chemical
potential. Also we develop simple procedure, which allows to
increase efficiency of the SSE algorithm in the general case.

\section{ The algorithm}
During the construction of the algorithm we follow the ideas of the
work \cite{Sand3}, therefore we do not describe SSE method in
details but briefly outline it.

Let us to consider Heisenberg model in the case of arbitrary spin
$S$, in the presence of external longitudinal field $h$
\begin{equation}
\hat{H} = \label{Hheis} \pm
J\sum_{<i,j>}{{\bf{S}}_{i}{\bf{S}}_{j}}-h\sum_{i}{S^{z}_{i}},
\end{equation}
and Bose Hubbard model with interaction
\begin{eqnarray}
\label{Hhub} \hat{H} =
-t\sum_{<i,j>}{(b^{+}_{i}b_{j}+b_{i}b^{+}_{j})}+V\sum_{<i,j>}{n_{i}n_{j}}
\nonumber\\+U\sum_{i}{n^{2}_{i}} -\mu\sum_{i}{n_{i}},
\end{eqnarray}
where $\langle i,j\rangle$ denotes summation over the pairs of
nearest-neighbor sites. Following to the ideas of the SSE method,
we rewrite the Hamiltonians~(\ref{Hheis},\ref{Hhub}) as a sum over
diagonal and off-diagonal bond operators
\begin{equation}
 \hat{H}= -J\sum_{<i,j>}(\hat{H}^{(d)}_{ij}\mp
 \hat{H}^{(n)}_{ij}),
\end{equation}
where minus corresponds to antiferromagnet, plus corresponds to
ferromagnet and Hubbard model (for Hubbard model  $J$ corresponds
to $t$). In the case of the Heisenberg model the operators are
\begin{eqnarray}
\label{Heis} \hat{H}^{(d)}_{ij}&=&C \mp
S^z_{i}S^z_{j}+\frac{h}{2J}(S^z_{i}+S^z_{j})\\
\hat{H}^{(n)}_{ij}&=&\frac12(S^+_{i}S^-_{j}+S^-_{i}S^+_{j}),\nonumber
\end{eqnarray}
and, correspondingly, in the case of the Bose Hubbard model the
operators are
\begin{eqnarray}
\label{Hub}
\hat{H}^{(d)}_{ij}&=&C-\frac{V}{t}n_{i}n_{j}-\frac{U}{2t}(n^2_{i}+n^2_{j})
+\frac{\mu}{2t}(n_{i}+n_{j})\\
\hat{H}^{(n)}_{ij}&=&b^{+}_{i}b_{j}+b_{i}b^{+}_{j}.\nonumber
\end{eqnarray}
One should guarantee non-negativity of all matrix elements of the
operators~(\ref{Heis},\ref{Hub})  by appropriate choosing of
constants C.

The SSE algorithm is based on the series expansion of the partition
function $Z$ with respect to inverse temperature $\beta$ powers. To
simplify Monte Carlo simulation, Sandvik et al. \cite{Sand2,Sand3}
propose to introduce unit operators $\hat{I}$ and cut off the
expansion at $n=L$ power. It should be point out, that unit
operators can be distributed in different ways. So we obtain the
formula for the partition function
\begin{equation}
Z=\sum_{\alpha}\sum_{\{S_L\}}\frac{(J\beta)^n(L-n)!}{L!}\langle
\alpha|\prod_{k=1}^L \hat{H}^{(\gamma)}_{k}|\alpha\rangle,
\end{equation}
where $\gamma$ denotes the operator type  - unit, diagonal,
non-diagonal,  $S_{L}$ is a sequence of operator indices and $n$ is
the number of non-unit operators in $S_{L}$ .

\begin{figure}
\begin{picture}(100,100)
\put(0,10){{\sl c)}} \put(20,10){\line(1,0){50}} \put(0,50){{\sl
b)}} \put(20,50){\line(1,0){50}} \put(0,90){{\sl a)}}
\put(20,90){\line(1,0){50}}\put(30,100){\circle{10}}\put(60,100){\circle*{10}}
\put(30,80){\circle{10}}\put(60,80){\circle*{10}}
\put(20,55){\circle*{5}}\put(27,55){\circle*{5}}\put(34,55){\circle{5}}
\put(20,45){\circle*{5}}\put(27,45){\circle*{5}}\put(34,45){\circle{5}}
\put(70,55){\circle*{5}}\put(63,55){\circle{5}}\put(56,55){\circle{5}}
\put(70,45){\circle*{5}}\put(63,45){\circle{5}}\put(56,45){\circle{5}}
\put(30,20){\oval(15,10)}\put(30,0){\oval(15,10)}\put(27,17){\bf{2}}\put(27,-3){\bf{2}}
\put(60,20){\oval(15,10)}\put(60,0){\oval(15,10)}\put(57,17){\bf{1}}\put(57,-3){\bf{1}}
\end{picture}
\caption{\small An example of different vertices. {\sl a)} In the
case of  hard-core bosons or $S=1/2$ Heisenberg model. {\sl b)} In
the case of the spin-split representation for the $S=3/2$
Heisenberg model. {\sl c)} Vertex {\sl b)}, in the filling number
representation.  $"1"$ is identified with the spin projection
$S^{z}=-1/2$, $"2"$ is identified with the spin projection
$S^{z}=1/2$. For the Bose Hubbard model one can identify $"1"$ with
one boson per site, $"2"$ with two bosons per site.} \label{vert}
\end{figure}
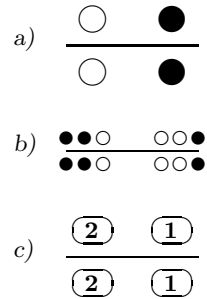

The Monte Carlo simulation is carried out with diagonal and loop
updates. The simulation starts with an arbitrary state $|\alpha>$
and operator string $S_{L}$ containing  only unit operators. During
the diagonal update one attempts to interchange diagonal and unit
operators with the probabilities
\begin{eqnarray}
P(\hat{I}\rightarrow\hat{H}^{(d)}_{ij})=\frac{ J N\beta\langle
\alpha(p)|\hat{H}^{(d)}_{ij}|\alpha(p)\rangle}{L-n}\\
P(\hat{H}^{(d)}_{ij}\rightarrow\hat{I})=\frac{L-n+1}{J
N\beta\langle \alpha(p)|\hat{H}^{(d)}_{ij}|\alpha(p)\rangle
}\nonumber,
\end{eqnarray}
where $|\alpha(p)\rangle$ is the system state after $p$ operators
been applied to it, $N$ is a number of bonds. Note that diagonal
update changes the expansion power $n$ by $\pm1$.

In the stage of loop update interchanging of diagonal and
non-diagonal operators is carried out with the fixed expansion
power $n$. At the same time system state $|\alpha>$ can be changed.

In the case of spin $S=1/2$ loop update is executed  in the
following way. Non-unit operators can be represented as vertices
with four legs (Fig.~\ref{vert}{\sl a)}). One of the $n$ vertices
is selected and one of its four legs is selected at random. After
that exit leg of the vertex is selected according with appropriate
probabilities and the spins at both the entrance and exit legs are
flipped. Note that the exit leg uniquely points to the entrance leg
of the next vertex. The loop is constructed such way until it
closed.

At  $S>1/2$ spin-split representation of spin operators is widely
used (Fig.~\ref{vert}{\sl b)}). In this case vertex contains
$4(2S+1)$ variables, which can take the value $\pm1$. During the
construction of the loop spins at the entrance and the exit legs
are flipped. But now loop propagates through the vertices with
$4(2S+1)$ legs. And therefore a number of possible loop pathes
increases rapidly with spin increase.

SSE algorithm allows to refuse spin-split representation and to
apply filling number representation which is applicable both for
Heisenberg and Bose Hubbard model. In order to do it we use well
known expressions for the matrix elements of corresponding
operators
\begin{eqnarray} \langle s|S^{+}|s-1 \rangle &=&\langle
s-1|S^{-}|s \rangle = \sqrt{(S+s)(S-s+1)} \nonumber\\ \langle
n|b^{+}|n-1 \rangle &=&\langle n-1|b|n \rangle = \sqrt{n}.
\end{eqnarray}
Now vertex has only four legs at arbitrary spin or at arbitrary
maximum filling number for bosons (Fig.~\ref{vert}{\sl a)}).
However, variables connected with legs take values $-S,...,S$ for
spins or $0,...,n_{max}$ for bosons. Therefore during the
construction of the loop we cannot use only flip of states at
entrance and exit legs. So we introduce increasing and decreasing
processes. To avoid discontinuous loop pathes during the
construction of loops we use a simple rule: if state at the exit
leg is decreased (increased) then at the entrance leg of the next
vertex decreasing (increasing) process will be chosen.

\section{Optimization of the algorithm}
Recently Sandvik and Sylju{\aa}sen \cite{Sand3} showed that in
order to fulfil detailed balance for loop update one should to
solve the set of equations
\begin{equation}
\label{Sys}
 W_{i}=\sum_{j}a_{ij},
\end{equation}
where $W_{i}$ are matrix elements of the operators
~(\ref{Heis},\ref{Hub}), and $a_{ij}$ are all allowed processes.
For example $a_{ii}$ denotes bounce process, which does not change
matrix element $W_{i}$, and $a_{ij}$ denotes process which
transforms $W_{i}$ to $W_{j}$.  It should be point out that all
$a_{ij}$ must be nonnegative and because of detailed balance
principle $a_{ij}=a_{ji}$. From $a_{ij}$ one can obtain
probabilities of all processes $P(W_{i}\rightarrow W_{j})=
a_{ij}/W_{i}$ and correspondingly $P(W_{j}\rightarrow W_{i})=
a_{ij}/W_{j}$.

\begin{figure}
\begin{center}
\resizebox{75mm}{!}{\includegraphics{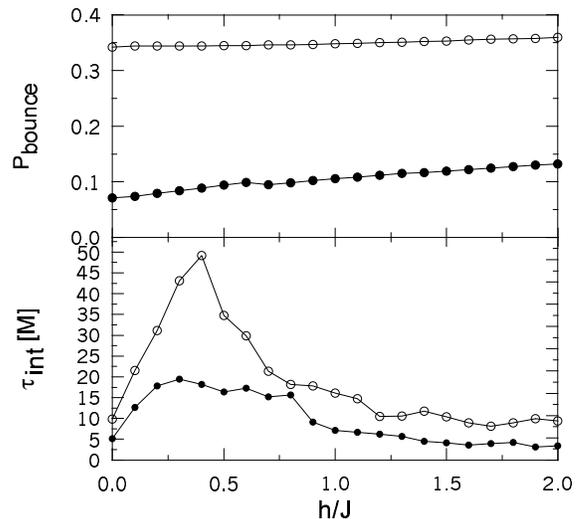}}
\end{center}
\caption{\small Upper plot - bounce probabilities vs external field
in the $S=5/2$ antiferromagnetic Heisenberg model at  $N_{s}=16$
and $\beta=10$. Coupling constant $J=1.0$. Dark circles correspond
to the optimized algorithm, open circles correspond to the
heat-bath algorithm. Lower plot - integrated autocorrelation times
for the magnetization vs external field in the cases of optimized
and heat-bath algorithms. } \label{opt}
\end{figure}

We found that in the case of arbitrary spin, set of all processes
$\{a_{ij}\}$ is decomposed into closed groups containing one, three
and six non-bounce processes. The group with one non-bounce process
is described by equations set with two equations, and groups with
three and six non-bounce processes are described by equations sets
with three and four equations correspondingly. So the equations
set~(\ref{Sys}) is decomposed into sets consisting of two, three
and four equations. Relations between number of various groups is
different at different values of spin. For example in the case of
$S=1/2$ there are only groups containing three non-bounce
processes. However at $S=1$  groups containing three and six
non-bounce processes appear. And number of such groups grows with
increase of spin until spin value becomes $S=5/2$. At $S=5/2$ part
of groups with one non-bounce process is $4/15$, with three
non-bounce processes is $3/15$ and with six non-bounce processes is
$8/15$. At $S>5/2$  the relations between number of groups are the
same as for $S=5/2$.

It is obvious that there is particular non-negative  solution of
the equations set~(\ref{Sys}). It is so-called heat-bath solution.
\begin{equation}
a_{ij} = \frac{W_{i}W_{j}}{\sum_{k}W_{k}}.
\end{equation}
In the denominator sum is over all matrix elements belonging to the
group. Unfortunately heat-bath solution gives rise to the
inefficient algorithm since all bounce processes $a_{ii}$ are
nonzero. In order to increase efficiency of algorithm, one should
to exclude bounce processes. Let us to do it for different types of
groups.

In the case of the group with one non-bounce process corresponding
set of equations is
\begin{eqnarray}
\label{Sys2} W_{1}&=&a_{11}+a_{12} \\
W_{2}&=&a_{22}+a_{21}.\nonumber
\end{eqnarray}
So we can always exclude one of bounce processes by choosing
$a_{12}=W_{2},a_{11}=W_{1}-W_{2},a_{22}=0$ if $W_{1}>W_{2}$ and
$a_{12}=W_{1},a_{22}=W_{2}-W_{1},a_{11}=0$ otherwise. It is obvious
that if $W_{1}=W_{2}$ bounce processes are absent at all.

\begin{figure}
\begin{center}
\resizebox{85mm}{!}{\includegraphics{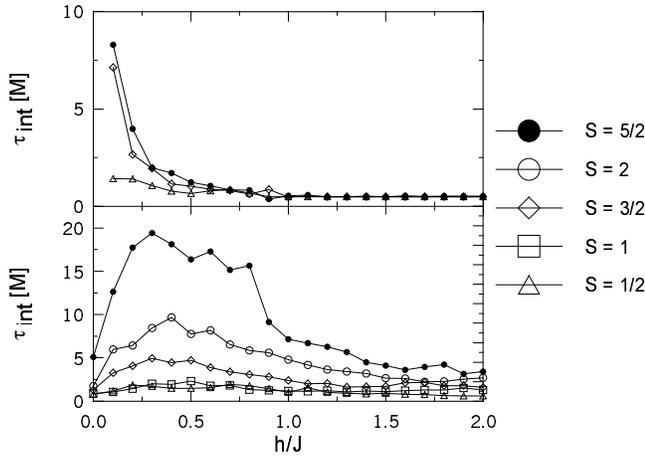}}
\end{center}
\caption{\small Integrated autocorrelation times for the
magnetization and energy vs external field in ferromagnetic (upper
plot) and antiferromagnetic (lower plot)  Heisenberg model with
different spin $S$ at $N_{s}=16$ and $\beta=10$. Coupling constant
is $J=1.0$.} \label{heis}
\end{figure}

Sandvik and Sylju{\aa}sen for  $S=1/2$ Heisenberg model have
analysed  analytically groups with three non-bounce processes
\cite{Sand3}, which are described by the equations set
\begin{eqnarray}
\label{Sys3}
 W_{1}&=&a_{11}+a_{12}+a_{13} \nonumber\\
W_{2}&=&a_{22}+a_{21}+a_{23}\\
W_{3}&=&a_{33}+a_{31}+a_{32}.\nonumber
\end{eqnarray}
They  proposed different solutions of the equations
set~(\ref{Sys3}) for various parameters of the model. It should be
point out, that some solutions contain two bounce processes. At the
same time for the case of arbitrary spin one cannot analytically
analyse all allowed processes and obtain corresponding
probabilities because number of processes grows rapidly as spin
increase.

We considered the equations set~(\ref{Sys3}) in general and
concluded that only one bounce process is needed at any $W_{i}$.
And there is no need to solve equations set~(\ref{Sys3})
analytically, but it is possible to use simple procedure for
obtaining non-negative solution of Eq.~(\ref{Sys3}).

First we demand all bounce processes $a_{ii}$ to be zero. Then
solution of Eq.~(\ref{Sys3}) takes the form
\begin{eqnarray}
\label{Sol3} a_{12}&=&\frac{W_{1}+W_{2}-W_{3}}{2} \nonumber\\
a_{13}&=&\frac{W_{1}+W_{3}-W_{2}}{2}\\
a_{23}&=&\frac{W_{2}+W_{3}-W_{1}}{2}.\nonumber
\end{eqnarray}
(We take into account that $a_{ij}=a_{ji}$.) If one of $a_{ij}$ is
negative then two others are certainly positive. So we need only
one bounce process. Let $a_{12}<0$ to be negative, then we should
introduce  bounce $a_{33}$ in a such way that $a_{12}$ becomes
positive and $a_{13},a_{23}$ do not change the sign. Let
$W_{1}>W_{2}$, then by choosing $a_{33}=W_{3}-W_{1}-W_{2}/\delta$
we get new solution of Eq.~(\ref{Sys3})
\begin{eqnarray}
\label{Sol31}
 a_{12}&=&\frac{W_{2}}{2}(1-\frac{1}{\delta})
\nonumber\\ a_{13}&=&W_{1}+ \frac{W_{2}}{2}(\frac{1}{\delta}-1)\\
a_{23}&=&\frac{W_{2}}{2}(1+\frac{1}{\delta}).\nonumber
\end{eqnarray}
It is obvious that at any $\delta>1$ solution~(\ref{Sol31}) is
positive. If $W_{2}>W_{1}$ one should to interchange $W_{1}$ by
$W_{2}$ in~(\ref{Sol31}). It should be point out that at $\delta=1$
solution~(\ref{Sol31}) coincides with some solutions proposed in
Ref.~\onlinecite{Sand3}. We do not assert that our solution is most
effective, but given procedure is universal and it can be applied
at arbitrary spin.

\begin{figure}
\begin{center}
\resizebox{85mm}{!}{\includegraphics{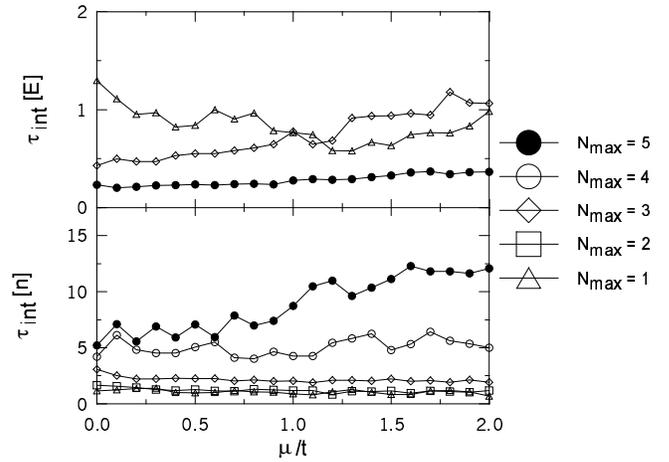}}
\end{center}
\caption{\small Integrated autocorrelation times for the mean
number of bosons and energy vs chemical potential in the Bose
Hubbard model with different maximum site filling  at $N_{s}=16$
and $\beta=10$. Hopping constant is $t=1.0$, $U=0.5$, $V=0.5$.}
\label{hub}
\end{figure}

The groups with six non-bounce processes are described by the
equations set
\begin{eqnarray}
\label{Sys4} W_{1}&=&a_{11}+a_{12}+a_{13}+a_{14} \nonumber\\
W_{2}&=&a_{22}+a_{21}+a_{23}+a_{24}\\
W_{3}&=&a_{33}+a_{31}+a_{32}+a_{34}\nonumber\\
W_{4}&=&a_{44}+a_{41}+a_{42}+a_{43}.\nonumber
\end{eqnarray}
As well in the case of group with three non-bounce processes, we
demand $a_{ii}=0$ and take into account $a_{ij}=a_{ji}$. Then we
obtain the equations set with four equations and six variables,
i.e. we have two free parameters. Let us assume
$a_{23}=a_{34}=a_{13}$, then we obtain solution of Eq.~(\ref{Sys4})
\begin{eqnarray}
\label{Sol4} a_{12}&=&\frac{W_{1}+W_{2}-W_{4}}{2}-\frac{W_{3}}{6}
\nonumber\\ a_{13}&=&\frac{W_{3}}{3}\\
a_{14}&=&\frac{W_{1}+W_{4}-W_{2}}{2}-\frac{W_{3}}{6}\nonumber\\
a_{24}&=&\frac{W_{2}+W_{4}-W_{1}}{2}-\frac{W_{3}}{6}.\nonumber
\end{eqnarray}
We can guarantee positivity of terms like $(W_{1}+W_{2}-W_{4})/2$,
by using procedure which we apply for the equations set with three
equations. Thus we introduce one bounce process. After that we
obtain expressions like $a-W_{3}/6$ with positive $a$. If latter
expression is negative one can add process
$a_{33}=W_{3}(1-1/\delta_{2})$. And we can provide positivity of
solution~(\ref{Sol4}) by choosing $\delta_{2}$ sufficiently large.

\section{Test calculations}

SSE algorithm is universal in any dimension. With increase of
dimension extra bonds arise, but ideas of loop construction remain
the same. Therefore we test the proposed scheme on 1D systems.

We calculate magnetization $M$ for Heisenberg model, a mean number
of bosons  $N_{b}$ for Bose Hubbard model, and energy for both
models. We use well-known estimators \cite{Sand3}
\begin{eqnarray}
\label{quant} E&=& -\frac{\langle n\rangle}{\beta}\nonumber\\
M&=&\frac{1}{N_{s}}\sum^{N_{s}}_{i=1}\langle S^{z}_{i}\rangle\\
N_{b}&=&\frac{1}{N_{s}}\sum^{N_{s}}_{i=1}\langle n_{i}\rangle,
\nonumber
\end{eqnarray}
where $n$ is a number of non-unit operators in operator string and
$N_{s}$ is a number of sites. We have checked our results with
exact diagonalization and have found  that relative deviation our
results from exact is less then $10^{-3}-10^{-4}$.

It is well-known that integrated autocorrelation times is a
quantitative measure of  efficiency of a Monte Carlo sampling. We
calculate autocorrelation times using bining method, which
described in Ref.\onlinecite{Evertz}

First of all it is interesting to analyse influence of bounce
processes on  efficiency of the algorithm. To this end we calculate
for Heisenberg antiferromagnet integrated autocorrelation times for
magnetization by using heat-bath solution and optimized algorithm
described in the previous section. We consider spin $S=5/2$ because
at this value all types of groups are present and relations between
number of groups do not change with further spin increase. As shown
in Fig.~(\ref{opt}), in the case of the optimized algorithm bounce
probabilities are less then in the case of heat-bath solution.
Accordingly autocorrelation times are less for the optimized
algorithm. For other calculations reported here the optimized
algorithm has been used.

Fig.~(\ref{heis}) shows autocorrelation times for  magnetization
versus external field for ferromagnetic (upper plot) and
antiferromagnetic (lower plot) Heisenberg model with different spin
$S$. Calculations have been  done for the chain with $N_{s}=16$
sites at $\beta=10$.

\begin{figure}
\begin{center}
\resizebox{75mm}{!}{\includegraphics{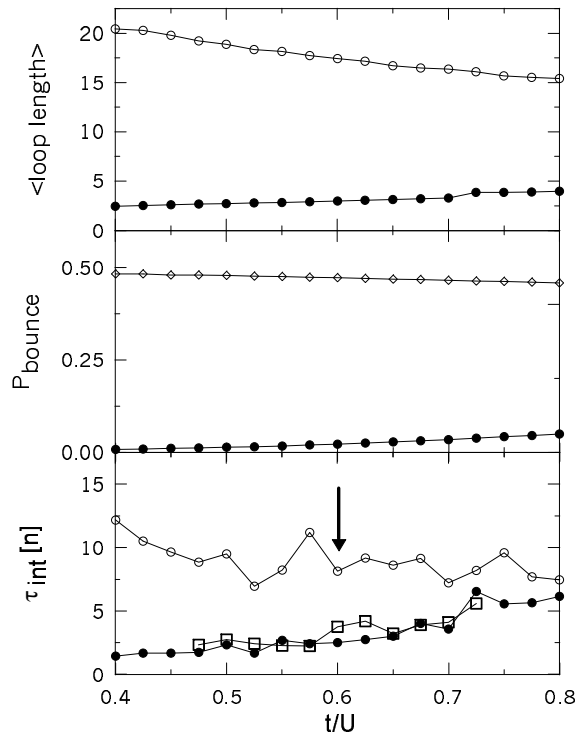}}
\end{center}
\caption{\small Lower plot - integrated autocorrelation times for
the mean number of bosons  in the Bose Hubbard model at $N_{s}=16$
and $\beta=10$, $N_{max}=5$. Hopping constant is $t=1.0$, $V=0.0$,
$\mu=U$. Dark circles correspond to the optimized algorithm, open
circles correspond to the heat-bath algorithm, squares correspond
to chain with $N_{s}=50$.  Middle plot - bounce probabilities for
optimized and heat-bath algorithms. Upper plot - mean length of
loops in units of $\langle n \rangle$ for optimized and heat-bath
algorithms accordingly. The arrow points to the critical point
"Mott insulator-superfluid" } \label{crit}
\end{figure}

One can see some increase of autocorrelation times with spin
increase for the antiferromagnet chain. However it is difficult to
compare efficiency of the algorithm at fixed temperature and
different spin. Mean number of non-unit operators can be roughly
estimated as $N_{s}J\beta S^{2}$. It is clear that this value grows
rapidly with the spin $S$ increase. We observe that the mean number
of non-unit operators $\langle n\rangle\sim100$ at $\beta=10$ and
$N_{s}=16$ in the case of spin $S=1/2$ whereas in the case of spin
$S=5/2$ at the same conditions $\langle n\rangle\sim 2000$. Thus
simulation of $S=5/2$ Heisenberg antiferromagnet at $\beta=10$ is
as hard as simulation of $S=1/2$ Heisenberg antiferromagnet at
$\beta\sim 100$. Hence the origin of autocorrelation time increase
is clear and, with the other hand, it is obvious that the algorithm
is very efficient. It should be point out that the algorithm works
efficiently in wide range of external fields.

For the ferromagnet chain we obtain good autocorrelation times for
magnetization in wide range of external fields with the exception
of zero field. At zero field autocorrelation times for
magnetization become very large (we do not show corresponding
points at Fig.~(\ref{heis})). It is a known sequence of degeneracy
states with spins up and spins down.

Also we done calculations for Bose Hubbard model with interaction.
As seen from Fig.~(\ref{hub}) autocorrelation times for energy is
order of unity. Autocorrelation times for mean number of bosons
grow with maximum filling number $N_{max}$ increase. Note that we
can use any maximum filling number $N_{max}$, and for large class
of problems the value $N_{max}\sim 5...10$ is  quite enough.

Investigation of many-body quantum system behavior near the
critical points is  one of interesting problem in  condenced matter
physics. Kawashima et al. have tested SSE directed loop algorithm
for 3D system and failed to obtain estimates for the observables
near the critical point \cite{Smak}. It is well known that 1D Bose
Hubbard model experiences superfluid-insulator transition at
$(t/U)_{c} = 0.608$, $V=0$, $\mu = U$ \cite{Kash2}. We calculate
autocorrelation times near the critical point for $N_{s}=16,
N_{s}=50$ chains at $\beta=10$, $N_{max}=5$. As seen from
Fig.~(\ref{crit}) autocorrelation times for both optimized and
heat-bath algorithms are quite reasonable. But bounce probabilities
in the case of heat-bath algorithm are very large and exceed bounce
probabilities in the case of optimized algorithm by order of
magnitude. Large bounce probabilities give rise to enormous loops,
which walk around system many times until closed. Construction of
such big loops takes a lot of time and simulation becomes
inefficient. So we can conclude that SSE algorithm allows to
perform simulations near the critical point (at least near
superfluid-insulator transition in 1D), however it is desirable to
exclude bounce processes.

\section{Summary}
In conclusion it should be emphasized that the algorithm introduced
here allows to explore  Heisenberg model with arbitrary spin and
Bose Hubbard model with interaction. With the help of filling
number representation we create the unified  code for both models.
Note that from algorithmic point of view differences between the
models arise only at stage of matrix elements calculation.

We  propose universal procedure for excluding bounce processes. It
has been obtained that for groups with one and three processes only
one bounce is needed and in the case of group with six processes
maximum two bounces are needed. We found that relations between
number of various groups are different up to spin $S=5/2$ (maximum
filling number $N_{max}=5$). After spin $S=5/2$ the relations do
not change.

Calculations of integrated autocorrelation times demonstrate
increase  efficiency of the algorithm under bounce processes
excluding.  We show that the proposed algorithm works   in wide
range of external fields both for Heisenberg model with arbitrary
spin $S$ and for Bose Hubbard model with interaction. Also we found
that the algorithm is efficient near the superfluid-insulator
transition.

We are grateful to I. A. Rudnev for support. We acknowledge
financial support from RFBR under Grant No. 03-02-16979.


\begin{thebibliography}{99}

\bibitem{Hirch} J.E. Hirch, R.L. Sugar, D. J. Scalapino, R. Blankenbecler, Phys. Rev. B
 {\bf 26}, 5033 (1982).

\bibitem{Scal} G. G. Batrouni, R. T. Scalettar, Phys. Rev. B {\bf 46}, 9051 (1992).

\bibitem{Evertz} N. Kawashima and J. E. Gubernatis, H. G. Evertz,
Phys. Rev. B {\bf 50}, 136 (1994).

\bibitem{Wiese} B. B. Beard and U.J. Wiese,
 Phys. Rev. Lett. {\bf 77}, 5130 (1996).

\bibitem{Kaw} H. Onishi, M. Nishino, N. Kawashima and S. Miyashita,
J. Phys. Soc. Jpn. {\bf 68} 2547 (1999), cond-mat/9903375.

\bibitem{Kash} V. A. Kashurnikov, N. V. Prokof'ev, B. V. Svistunov, and M. Troyer,
Phys. Rev. B {\bf 59}, 1162 (1999).

\bibitem{Sand1} A. W. Sandvik, R. R. P. Singh, D. K. Campbell,
Phys. Rev. B {\bf 56}, 9051 (1997).

\bibitem{Sand2} A. W. Sandvik, Phys. Rev. B {\bf 59}, R14157 (1999).

\bibitem{Sand3} O. F. Sylju{\aa}sen, A. W. Sandvik, Phys. Rev. E
{\bf66}, 046701 (2002).

\bibitem{sp} A.W. Sandwik, Phys. Rev. B {\bf 66}, 024418 (2002).\\ S.
Wessel, M. Olshanii, and S. Haas, Phys. Rev. Lett. {\bf 87}, 206407
(2001).\\ S. Yunoki, Phys. Rev. B {\bf 65}, 092402 (2002).

\bibitem{bos} A. Dorneich, W. Hanke, E. Arrigoni, M. Troyer, and S.C.Zhang,
Phys. Rev. Lett. {\bf 88}, 057003 (2002).\\F. Hebert, G. G.
Batrouni, R. T. Scalettar, G. Schmid, M. Troyer, and A. Dorneich,
Phys. Rev. B {\bf 65}, 014513 (2002).\\ G. Schmid, S. Todo, M.
Troyer, and A. Dorneich, Phys. Rev. Lett. {\bf 88}, 167208 (2002).

\bibitem{ferm}R. T. Clay, S. Mazumdar, and D. K. Campbell, Phys. Rev. Lett. {\bf 86},
4084 (2001).\\  P. Sengupta, A. W. Sandvik, and D. K. Campbell,
Phys. Rev B {\bf 65}, 155113 (2002).

\bibitem{Todo} S. Todo and K. Kato Phys. Rev. Lett. {\bf 87}. \\
K. Harada, M. Troyer and N. Kawashima, J. Phys. Soc. Jpn. {\bf67}
1130 (1998), cond-mat/9712292.

\bibitem{spin1} S. Bergkvist, P. Henelius, and A. Rosengren, Phys.
Rev. B {\bf66} 134407 (2002).

\bibitem{sps} P. Henelius, P. Fr\"{o}brich, P. J. Kuntz, C. Timm, and P. J. Jensen
Phys. Rev. B {\bf66}, 094407 (2002).

\bibitem{Smak} J. Smakov, K. Harada and N. Kawashima,
cond-mat/0301416.

\bibitem{Kash2} V. A. Kashurnikov and B. V. Svistunov, Phys. Rev. B
{\bf 53} 11776 (1996).

\end{thebibliography}
\end{document}